\documentclass[a4paper, 11pt]{article}
\usepackage{amsmath,graphicx,subfigure}
\usepackage[small]{caption2}

\setcaptionwidth{0.9\textwidth}
\newcommand{\coloneq}{\; \colon \mspace{-12.0mu} =}

\begin{document}

\begin{center}
{\Large \bf Stabilising the Blue Phases} \\[3mm]
G. P. Alexander and J. M. Yeomans \\ {\it Rudolf Peierls Centre for Theoretical Physics, University of Oxford, 1 Keble Road, Oxford, OX1 3NP.} \\[2mm]
\today
\end{center}

\begin{abstract}
We present an investigation of the phase diagram of cholesteric liquid crystals within the framework of Landau - de Gennes theory. The free energy is modified to incorporate all three Frank elastic constants and to allow for a temperature dependent pitch in the cholesteric phase. It is found that the region of stability of the cubic blue phases depends significantly on the value of the elastic constants, being reduced when the bend elastic constant is larger than splay and when twist is smaller than the other two. Most dramatically we find a large increase in the region of stability of blue phase I, and a qualitative change in the phase diagram, in a system where the cholesteric phase displays helix inversion.

\vspace{5 mm}
\end{abstract}

\numberwithin{equation}{section}

\section{Introduction}
\label{sec:intro}

Liquid crystals are anisotropic fluids typically composed of long, thin, rod-like molecules. They display long-range correlations in molecular orientation and show large length scale deformations to even relatively weak external perturbations. As a result they may be well described at the continuum level by a vector field called the director \cite{degennes,stewart}, which represents the average local molecular alignment.

A topic of long standing interest is the role of chirality in liquid crystals. The addition of small quantities of chiral dopant to a liquid crystal results in the appearance of a periodic structure in the molecular orientation with length scales typically in the optical range. The most common form for this periodic structure is a helical arrangement, known as the cholesteric phase, where the molecules display a natural twist along one direction. However, one of the most interesting features of chirality is that it also allows for more complicated structures. Although the cholesteric is always the thermodynamically stable phase at sufficiently low temperatures, it is found experimentally that upon cooling from the isotropic fluid, systems of high chirality display a series of first order phase transitions to brightly coloured `blue phases' (BPs) before the cholesteric is reached. There are as many as three thermodynamically distinct blue phases in the absence of external fields, all of which appear only in a narrow temperature range (typically $\sim 1 K$) just below the clearing point. Two of them, BPI and BPII, display selective Bragg reflections in the visible range which can be indexed by cubic space groups, $O^{8-}(I4_132)$ and $O^2(P4_232)$ respectively, while the third, BPIII, is amorphous.

In the early and mid 1980's a series of important theoretical works showed how the cubic blue phases could be understood within both of the two principal theoretical continuum models of liquid crystals; the Oseen-Frank theory for the director field \cite{meiboom}, and the Landau - de Gennes theory based on the $\mathbf{Q}$-tensor order parameter \cite{ghsa,ghsb} (a review of both approaches is given in \cite{wright}). In both cases the picture which emerged was one of frustration between competing effects. From the director field point of view the locally prefered structure is one of `{\it double twist}', but this is found to be incompatible with global topological requirements. Therefore the cubic blue phases emerge as a regular array of double twist cylinders separated by a network of disclination lines. Complementing this the Landau - de Gennes theory picture of blue phases is that of a linear combination of biaxial helices chosen to optimise the competing bulk and gradient free energies.

The theoretical models were successful in accounting for the occurance, symmetry and general properties of the cubic blue phases, even assisting in the determination of specific space groups. However, the approximations necessary to make the analytic calculations feasible meant that quantitative comparison with experiment was largely not possible. Furthermore, the theories have a certain inflexibility since they retain a bare minimum of parameters and in this sense provide a `one size fits all' description of the blue phases. This may have been adequate twenty years ago, but as the field has developed its shortcomings have become more apparent. One example of this is the insensitivity of the Landau - de Gennes theory of the blue phases to the value of the twist elastic constant. In contrast, the Frank director theory predicts that the twist elastic constant plays a central role \cite{meiboom} and furthermore, it has been observed experimentally that the stability of the cubic blue phases depends as much on the value of the twist elastic constant as it does on the chirality \cite{miller}.

More recently there has been considerable interest in trying to manipulate the properties of the cubic blue phases by adding a variety of chemical dopants \cite{nakata,chanishvili,kikuchi,yoshizawa}, with the long term goal of utilising blue phases for device applications \cite{cao,kitzerow,hisakado,yokoyama}. One of the major obstacles to the use of blue phases in devices is their limited range of thermodynamic stability. However, Coles and Pivnenko have recently reported achieving a $40 K$ range of stability for BPI in a bi-mesogenic compound doped with a chiral additive \cite{coles}. Although some questions remain as to whether their blue phase is thermodynamically stable or just metastable, it seems clear that the current theoretical picture of the blue phases is insufficient to account for all of the reported observations. Therefore, in this work, we ask whether extending the traditional Landau - de Gennes theory beyond a one elastic constant approximation will allow for a better comparison with experiments and perhaps aid the search for more practically useful blue phases.

We first present a comparison of the Frank director and Landau - de Gennes theories, showing the relationships between them and discussing the limitations of both. Then we show how the Landau - de Gennes theory can be modified to account for three independent Frank elastic constants and a temperature dependent helical pitch in the cholesteric phase. The modified theory is treated in detail analytically for the cholesteric phase and it is shown that the theory allows for a description of the change in sense of the cholesteric helix upon decreasing temperature, which is observed in some systems \cite{degennes,huff,slaney}.  Next we present a numerical determination of the thermodynamic phase diagram of chiral liquid crystals for a range of physical parameters. To do this we introduce a new technique for determining the unit cell size of the blue phases, allowing for the first time, for a full minimisation of the free energy. Although the qualitative features remain unchanged when the elastic constants are varied, there is significant quantitative movement of the phase boundaries and the range of stability of the blue phases is brought to lower values of the chirality than has been reported previously. Our most striking result is that the range of stability of BPI is increased dramatically in systems where the cholesteric undergoes helical sense inversion.

\section{The Frank free energy and Landau - de Gennes theory}
\label{sec:cubicinvariants}

The most widely used description of liquid crystals is the Frank director field theory \cite{degennes,stewart}. The local orientation of the liquid crystal is described by a directionless unit vector field $\mathbf{n}(\mathbf{r})$, called the director. It is found experimentally that the director satisfies an equivalence relation $\mathbf{n} \sim -\mathbf{n}$, so that its configuration space is the real projective plane, $\mathbf{n}\in \mathbf{RP}^2$. The static properties of liquid crystals are well described by director field configurations which minimise the Frank free energy subject to appropriate boundary conditions
\begin{equation}
F = \tfrac{1}{V} \int d^3r \; \Bigl( \tfrac{1}{2} K_{11}^{\text{F}} \bigl( \nabla \cdot \mathbf{n} \bigr)^2 + \tfrac{1}{2} K_{22}^{\text{F}} \bigl( \mathbf{n} \cdot \nabla \times \mathbf{n} + q_0^{\text{F}} \bigr)^2 + \tfrac{1}{2} K_{33}^{\text{F}} \bigl( \mathbf{n} \times \nabla \times \mathbf{n} \bigr)^2 \Bigr) \; ,
\label{eq:frank}
\end{equation}
where $K_{11}^{\text{F}},K_{22}^{\text{F}}, \text{and}\; K_{33}^{\text{F}}$ are the Frank elastic constants, known as {\it splay}, {\it twist} and {\it bend} repectively. The parameter $q_0^{\text{F}}$ is nonzero only in chiral liquid crystals where it determines the pitch of the cholesteric helix. 

In discussions of the cholesteric blue phases the Frank free energy is supplemented by a {\it saddle-splay} term \cite{degennes,stewart}
\begin{equation}
\tfrac{1}{2} K_{24}^{\text{F}} \nabla \cdot \Bigl[ \bigl( \mathbf{n} \cdot \nabla \bigr) \mathbf{n} - \mathbf{n} \bigl( \nabla \cdot \mathbf{n} \bigr) \Bigr] \; .
\label{eq:saddlesplay}
\end{equation} 
In ordinary liquid crystals this term can be safely treated as a total divergence and integrated to a surface term, which is then discarded. However, in systems possessing defects, such as the blue phases, variations in the magnitude of the order become as relevant as variations in the direction of the order. The value of the Frank elastic constants depends on the magnitude of the order (in a manner which can be determined using the Landau - de Gennes theory) so that in such cases the saddle-splay invariant may not be treated as a total divergence and can play an important role in the energetics of systems with defects. Indeed it has been shown that, within the director field framework, the saddle-splay term is entirely responsible for the thermodynamic stability of the blue phases \cite{degennes,meiboom,wright}.

There are two principal drawbacks of the director field theory; first, the magnitude of the order has to be put in by hand, and second, the effect of biaxiality is not included. Both of these shortcomings are most severe in the vicinity of defects where it is necessary to adopt a more sophisticated framework in order to give a complete description. This is a particular problem for theories of the blue phases since they contain a regular lattice of defects. An alternative option is to use a traceless symmetric second-rank tensor, $\mathbf{Q}$, as a more general order parameter for liquid crystals. From a physical point of view it may be thought of as proportional to the anisotropic part of the magnetic susceptibility, or dielectric tensor. A phenomenological description of the energetics and phase transitions of liquid crystals can be provided by constructing a Landau theory using the $\mathbf{Q}$-tensor \cite{degennes}. This comprises a bulk contribution
\begin{equation}
F^{\text{bulk}} = \tfrac{1}{V} \int d^3r \;\Bigl( a\, \text{tr} (\mathbf{Q}^2) - b\, \text{tr} (\mathbf{Q}^3) + c \bigl( \text{tr} (\mathbf{Q}^2) \bigr)^2 \Bigr) \; ,
\label{eq:febulk}
\end{equation} 
and a gradient contribution, accounting for the energy cost associated with distortions of the order,
\begin{equation}
F^{\text{grad}} = \tfrac{1}{V} \int d^3r \;\Bigl( \tfrac{1}{4}\,L_1 \bigl( \nabla \times \mathbf{Q} + 2q_0 \mathbf{Q} \bigr)^2 + \tfrac{1}{4}\,L_2 \bigl( \nabla \cdot \mathbf{Q} \bigr)^2 \Bigr) \; .
\label{eq:fegrad}
\end{equation}
The bulk free energy describes a first order phase transition between the ordered and disordered fluids. The parameters $b$ and $c$ are positive constants and $a$ is a thermal scaling variable, which changes sign with decreasing temperature and has a linear dependence sufficiently close to the critical surface. The parameter $q_0$ defines the helical pitch. Since it is assumed that $q_0$ is independent of temperature it follows that so too is the cholesteric pitch. It is also clear that this theory, taken to second order in the derivatives of $\mathbf{Q}$, can only account for two independent Frank elastic constants; it has long been known that the constraint is that splay equals bend, $K_{11}^{\text{F}} = K_{33}^{\text{F}}$. In order to remove this degeneracy, and allow for a temperature dependent helical pitch, it is necessary to retain higher order terms in the gradient free energy. 

The construction of invariants contributing to the gradient free energy at higher than quadratic order has been considered previously by a number of researchers \cite{berreman, longa}. Here, for completeness, we list the possible invariants which can be formed at cubic order in the $\mathbf{Q}$-tensor, and at most quadratic order in gradients.
\begin{subequations}
\begin{gather}
\epsilon_{abc} Q_{ad} Q_{de} \nabla_b Q_{ce} \label{eq:cubicchiral} \; ,\\
\epsilon_{abc} Q_{ad} Q_{be} \nabla_d Q_{ce} \; ,\\
Q_{ab} \nabla_a Q_{bc} \nabla_d Q_{cd} \; ,\\
Q_{ab} \nabla_a Q_{cd} \nabla_b Q_{cd} \label{eq:cubicachiral} \; ,\\
Q_{ab} \nabla_a Q_{cd} \nabla_c Q_{bd} \; ,\\
Q_{ab} \nabla_c Q_{ab} \nabla_d Q_{cd} \; ,\\
Q_{ab} \nabla_c Q_{ac} \nabla_d Q_{bd} \; ,\\
Q_{ab} \nabla_c Q_{ad} \nabla_c Q_{bd} \label{eq:cubicachiralb} \; ,\\
Q_{ab} \nabla_c Q_{ad} \nabla_d Q_{bc} \; .
\end{gather}
\label{cubicinvariants}
\end{subequations}
We note that only six of the achiral invariants are linearly independent if, as is usual, we neglect total divergences, since we can construct the identity
\begin{multline}
Q_{ab} \nabla_c Q_{ad} \nabla_d Q_{bc} + Q_{ab} \nabla_a Q_{cd} \nabla_c Q_{bd} - Q_{ab} \nabla_c Q_{ac} \nabla_d Q_{bd} \\ 
- Q_{ab} \nabla_a Q_{bc} \nabla_d Q_{cd} = \nabla_c \Bigl( Q_{ab}Q_{ad}\nabla_d Q_{bc} - Q_{ab}Q_{ac}\nabla_d Q_{bd} \Bigr) \; .
\end{multline} 

Since there are now many more Landau - de Gennes invariants than there are Frank elastic constants, it is instructive to see how the two descriptions are related, which may be done by letting the $\mathbf{Q}$-tensor assume a uniaxial form, $Q_{ab} = S(3n_an_b-\delta_{ab})$, and identifying the director field with the eigenvector of the $\mathbf{Q}$-tensor corresponding to its maximal eigenvalue. A straightforward, but lengthy, calculation gives the following relations between the $\mathbf{Q}$-tensor invariants and the director field:
\begin{align}
\begin{split}
\bigl( \nabla \times \mathbf{Q} \bigr)^2 & = 18S^2 \bigl( \nabla_a n_b \bigr)^2 + 5 \bigl( \nabla_a S \bigr)^2 - 9S^2 \bigl( (\nabla_a n_b)(\nabla_b n_a) + (\mathbf{n} \times \nabla \times \mathbf{n})^2 \bigr)\\
& \quad - 3 \bigl( \mathbf{n}\cdot \nabla S \bigr)^2  - 6S\bigl( \nabla_a S\bigr) \bigl( 2\mathbf{n}\cdot \nabla n_a - n_a \nabla \cdot \mathbf{n} \bigr) \; ,
\end{split} \label{eq:cubicuniaxialstart} \\
\begin{split}
\bigl( \nabla \cdot \mathbf{Q} \bigr)^2 & = 9S^2 \bigl( (\nabla \cdot \mathbf{n})^2 + (\mathbf{n} \times \nabla \times \mathbf{n})^2 \bigr) + 3 \bigl( \mathbf{n}\cdot \nabla S \bigr)^2 + \bigl( \nabla_a S \bigr)^2 \\
& \quad  + 6S\bigl( \nabla_a S\bigr) \bigl( 2n_a \nabla \cdot \mathbf{n} - \mathbf{n}\cdot \nabla n_a \bigr) \; ,
\end{split} \\
\mathbf{Q} \cdot \nabla \times \mathbf{Q} & = 9S^2\; \mathbf{n} \cdot \nabla \times \mathbf{n} \; ,\\
\epsilon_{abc} Q_{ad} Q_{de} \nabla_b Q_{ce} & = 9S^3\; \mathbf{n}\cdot \nabla \times \mathbf{n} \; ,\\
\epsilon_{abc} Q_{ad} Q_{be} \nabla_d Q_{ce} & = 9S^3\; \mathbf{n}\cdot \nabla \times \mathbf{n} \; ,\\
\begin{split}
Q_{ab} \nabla_a Q_{bc} \nabla_d Q_{cd} & = 9S^3 \bigl( 2(\mathbf{n} \times \nabla \times \mathbf{n})^2 - (\nabla \cdot \mathbf{n})^2 \bigr) + S^3 \nabla_a (\mathbf{n} \cdot \nabla n_a - 2n_a \nabla \cdot \mathbf{n}) \\ & \quad + 9S (\mathbf{n} \cdot \nabla S )^2 - S(\nabla S)^2 \; ,
\end{split} \\
Q_{ab} \nabla_a Q_{cd} \nabla_b Q_{cd} & = 18S^3 \bigl( 3(\mathbf{n}\times \nabla \times \mathbf{n})^2 - (\nabla_an_b)^2 \bigr) + 18S (\mathbf{n} \cdot \nabla S)^2 - 6S(\nabla S)^2 \; ,\\
\begin{split}
Q_{ab} \nabla_a Q_{cd} \nabla_c Q_{bd} & = 9S^3 \bigl( 2(\mathbf{n} \times \nabla \times \mathbf{n})^2 - (\nabla_an_b)(\nabla_bn_a) \bigr) \\ & \quad - S^3 \nabla_a (2\mathbf{n}\cdot\nabla n_a - n_a \nabla \cdot \mathbf{n}) + 9S(\mathbf{n}\cdot\nabla S)^2 - S(\nabla S)^2 \; ,
\end{split} \\
Q_{ab} \nabla_c Q_{ab} \nabla_d Q_{cd} & = 18S(\mathbf{n}\cdot \nabla S)^2 - 6S(\nabla S)^2 - 6S^3 \nabla_a ( \mathbf{n} \cdot \nabla n_a + n_a \nabla \cdot \mathbf{n}) \; ,\\
\begin{split}
Q_{ab} \nabla_c Q_{ac} \nabla_d Q_{bd} & = 9S^3 \bigl( 2(\nabla \cdot \mathbf{n})^2 - (\mathbf{n} \times \nabla \times \mathbf{n})^2 \bigr) \\ & \quad - 2S^3 \nabla_a (\mathbf{n}\cdot \nabla n_a + 4n_a \nabla \cdot \mathbf{n}) + 9S(\mathbf{n} \cdot \nabla S)^2 - S(\nabla S)^2 \; ,
\end{split} \\
Q_{ab} \nabla_c Q_{ad} \nabla_c Q_{bd} & = 6S(\nabla S)^2 + 9S^3 (\nabla_an_b)^2 \; ,\\
\begin{split}
Q_{ab} \nabla_c Q_{ad} \nabla_d Q_{bc} & = 9S^3 \bigl( 2(\nabla_an_b)(\nabla_bn_a) - (\mathbf{n} \times \nabla \times \mathbf{n})^2 \bigr) \\ & \quad - 2S^3 \nabla_a ( 4 \mathbf{n}\cdot \nabla n_a + n_a \nabla \cdot \mathbf{n}) + 9S (\mathbf{n}\cdot \nabla S)^2 - S(\nabla S)^2 \; .
\end{split}
\label{eq:cubicuniaxialend}
\end{align}
Using the following identities for the director field
\begin{align}
(\nabla_a n_b)^2 & = (\nabla \cdot \mathbf{n})^2 + (\mathbf{n}\cdot\nabla\times\mathbf{n})^2 + (\mathbf{n}\times\nabla\times \mathbf{n})^2 + \nabla_a ( \mathbf{n}\cdot\nabla n_a -  n_a \nabla \cdot \mathbf{n}) \; ,\\
(\nabla_a n_b)(\nabla_b n_a) & = (\nabla \cdot \mathbf{n})^2 + \nabla_a (\mathbf{n}\cdot\nabla n_a -  n_a \nabla \cdot \mathbf{n}) \; ,
\label{directoridentities}
\end{align}
in Equations \eqref{eq:cubicuniaxialstart} to \eqref{eq:cubicuniaxialend} we can identify the Frank elastic constants in terms of the coefficients of the Landau - de Gennes theory as
\begin{align}
2 K_{11}^{\text{F}} & = 9S^2(L_{21}+L_{22})+9S^3(-L_{33}-2L_{34}-L_{35}+2L_{37}+L_{38}+2L_{39}) \; , \label{franksplay} \\
2 K_{22}^{\text{F}} & = 18S^2L_{21}+9S^3(-2L_{34}+L_{38}) \; , \label{franktwist} \\
2 K_{33}^{\text{F}} & = 9S^2(L_{21}+L_{22})+9S^3(2L_{33}+4L_{34}+2L_{35}-L_{37}+L_{38}-L_{39}) \; , \label{frankbend} \\
4 K_{22}^{\text{F}}q_0^{\text{F}} & = 36q_0S^2L_{21}+9S^3(L_{31}+L_{32}) \; . \label{frankpitch}
\end{align}

It is not realistic to study the effect of all possible cubic terms in detail and consequently we shall restrict our attention to a subset. We choose to focus on three invariants, \eqref{eq:cubicchiral}, \eqref{eq:cubicachiral} and \eqref{eq:cubicachiralb}. These are chosen for a number of reasons; first, it is clear that one should retain at least one chiral and one achiral invariant. Since there does not appear to be any substantial difference between the two chiral invariants we feel confident that no qualitative changes would result from retaining both terms. It is more difficult to choose representatives of the achiral invariants. The invariant \eqref{eq:cubicachiralb} is retained because uniquely amongst the achiral invariants it contributes equally to all three Frank elastic constants. This leaves only a choice of invariant that will distinguish between the Frank splay and bend elastic constants. We choose to use \eqref{eq:cubicachiral} partly because it has been used in previous work \cite{jung,toth}, partly because it gives the largest distinction between $K_{11}^{\text{F}}$ and $K_{33}^{\text{F}}$ and partly because we found that it has the largest contribution to the energetics of a single double twist cylinder.
 
To summarise, in the remainder of this paper we shall investigate the properties of the free energy
\begin{equation}
\begin{split}
F & = \tfrac{1}{V} \int d^3r \; \Bigl( \tfrac{\tau}{4} \, \text{tr}(\mathbf{Q}^2) - \sqrt{6} \, \text{tr}(\mathbf{Q}^3) + \bigl( \text{tr}(\mathbf{Q}^2) \bigr)^2 + \tfrac{\kappa^2}{4} \Bigl[ \bigl( \nabla \times \mathbf{Q} + \mathbf{Q} \bigr)^2 + \tfrac{L_{22}}{L_{21}} \bigl( \nabla \cdot \mathbf{Q} \bigr)^2 \\
& \qquad + \tfrac{L_{31}}{L_{21}} \mathbf{Q}^2 \cdot \nabla \times \mathbf{Q} + \tfrac{L_{34}}{L_{21}} Q_{ab}\nabla_a Q_{cd}\nabla_b Q_{cd} + \tfrac{L_{38}}{L_{21}} Q_{ab}\nabla_c Q_{ad}\nabla_c Q_{bd} \Bigr] \Bigr) \; .
\end{split}
\label{eq:ldgfreeenergy}
\end{equation}
We have made use of the well-known rescaling, $\mathbf{Q} \rightarrow (b/\sqrt{6}c) \mathbf{Q}$, to reveal the redundancy of the parameters $b$ and $c$ in the bulk free energy and shifted to dimensionless variables. The change of variables is $r \rightarrow r/2q_0, \, F \rightarrow b^4/(288q_0^3c^3)F$ and we define 
\begin{align}
L_{21} &\coloneq 6cL_1/b^2 \; , &  L_{22} &\coloneq 6cL_2/b^2 \; , \\ \kappa &\coloneq 2q_0 \sqrt{L_{21}} \; , &  \tau &\coloneq 6ac/b^2 \; .
\label{eq:deftaukappa}
\end{align}
We note that the parameter $\kappa$, which is called the chirality, is the same as that defined by Grebel {\it et. al.} \cite{ghsa}, but that our reduced temperature $\tau$ differs from their definition; the two being related according to $t_{\text{GHS}} = \tau + \kappa^2$, where $t_{\text{GHS}}$ is the reduced temperature used in \cite{ghsa}.

\section{Cholesteric Phase}
\label{sec:cholesteric}

We first investigate how including cubic order invariants in the gradient free energy, Equation \eqref{eq:ldgfreeenergy}, modifies the usual theory of the cholesteric helix. It will be assumed that the helical order parameter still takes the form dictated by the quadratic theory, namely
\begin{equation}
\mathbf{Q} = \frac{-Q_h}{\sqrt{6}} \begin{pmatrix} 2 & 0 & 0 \\ 0 & -1 & 0 \\ 0 & 0 & -1 \end{pmatrix} + \frac{Q_2}{\sqrt{2}} \begin{pmatrix} 0 & 0 & 0 \\ 0 & \cos (kx) & \sin (kx) \\  0 & \sin (kx) & -\cos (kx) \end{pmatrix} \; .
\label{eq:Qchol}
\end{equation}
With this form of $\mathbf{Q}$-tensor the expression for the free energy becomes \cite{comment1}
\begin{multline}
F_{\text{cholesteric}} = \tfrac{1}{4}(\tau + \kappa^2)(Q_h^2 + Q_2^2) + (Q_h^3 - 3Q_hQ_2^2) + (Q_h^2 + Q_2^2)^2 \\ + \tfrac{\kappa^2}{4} Q_2^2 \left[ (1 - \beta Q_h)k^2 - 2(1-\alpha Q_h)k \right] \; ,
\label{fechol}
\end{multline}
where, $\alpha \coloneq -L_{31}/(\sqrt{6}L_{21})$ and $\beta \coloneq (2L_{34}-L_{38})/(\sqrt{6}L_{21})$. Minimising the free energy leads to the set of Euler-Lagrange equations 
\begin{subequations}
\begin{gather}
k = \tfrac{1-\alpha Q_h}{1-\beta Q_h} \; , \label{kmin} \\
\tfrac{1}{2}(\tau + \kappa^2)Q_h + 3(Q_h^2 - Q_2^2) + 4Q_h(Q_h^2 + Q_2^2) + \tfrac{\kappa^2}{4} Q_2^2 \left[ -\beta k^2 + 2 \alpha k \right] = 0 \; ,\\
\tfrac{1}{2}(\tau + \kappa^2)Q_2 - 6Q_hQ_2 + 4Q_2(Q_h^2 + Q_2^2) + \tfrac{\kappa^2}{2} Q_2 \left[ (1 - \beta Q_h)k^2 - 2(1-\alpha Q_h)k \right]  = 0 \; .
\end{gather}
\end{subequations}
Using these, and assuming $Q_2 \not = 0$ as expected for a first order transition, we readily obtain
\begin{equation}
Q_2^2 = \tfrac{1}{8} \biggl( 12Q_h + \kappa^2 \tfrac{(1-\alpha Q_h)^2}{1-\beta Q_h} - (\tau + \kappa^2) \biggr) - Q_h^2 \; ,
\label{Q2zero}
\end{equation}
\begin{equation}
18Q_h^2 + \kappa^2Q_h\tfrac{(1-\alpha Q_h)^2}{1-\beta Q_h}  = Q_2^2 \biggl( 6-\tfrac{\kappa^2}{2} \Bigl[ 2\alpha -\beta \tfrac{1-\alpha Q_h}{1-\beta Q_h} \Bigr] \tfrac{1-\alpha Q_h}{1-\beta Q_h} \biggr) \; . \label{QhQ2} 
\end{equation}
In general these equations must be solved numerically; however, they can be easily solved when $\alpha = \beta = 0$, and we recover the usual results \cite{ghsa, wright, belyakov}
\begin{subequations}
\begin{align}
Q_h & = \tfrac{9-\kappa^2}{48} \Bigl( 1 + \sqrt{1-\tfrac{72\tau}{(9-\kappa^2)^2}} \Bigr) \; , \\
Q_2^2 & = \tfrac{27+\kappa^2}{24} Q_h - \tfrac{3\tau}{32} \; .
\end{align}
\end{subequations}

The main feature which arises from the inclusion of higher order terms is that the helical wavevector now depends on the temperature. As Equation \eqref{kmin} shows, the wavevector depends on the amplitude of the homogeneous part of the order parameter, and since the order parameter depends on the temperature, it follows that so does the cholesteric pitch. It should be noted that, even with such a simple modification of the free energy, we can already account for the experimentally observed inversion of the helical sense with decreasing temperature found in some cholesterics \cite{degennes, huff, slaney}. According to the present theory the cholesteric wavevector changes sign when $Q_h = 1/\alpha$. When this value is substituted into Equations \eqref{Q2zero} and \eqref{QhQ2} we obtain the following expression for the inversion temperature
\begin{equation}
\tau_{\text{HI}} = \tfrac{4}{\alpha^2}(3\alpha -8) - \kappa^2 \; .
\label{hitemp}
\end{equation} 
At the inversion itself we find $Q_2 = \sqrt{3}Q_h$ and
\begin{equation}
\mathbf{Q} = \frac{2}{\alpha \sqrt{6}} \begin{pmatrix} -1 & 0 & 0 \\ 0 & 2 & 0 \\ 0 & 0 & -1 \end{pmatrix} \; ,
\label{hiqtensor} 
\end{equation}
which corresponds as expected to a uniaxial nematic with director $\mathbf{n} = (0,1,0)$. Other orientations of the nematic within the $yz$-plane may be obtained by adding a constant phase shift to the $\sin$ and $\cos$ terms in $\mathbf{Q}$ in Equation \eqref{eq:Qchol}. It is clear that within this theory, as observed experimentally, the helical inversion is a smooth transition which does not involve any discontinuities in physical quantities, i.e., it is not a phase transition. 

The isotropic-cholesteric transition temperature is obtained by supplementing Equations \eqref{Q2zero} and \eqref{QhQ2} with the condition that the free energy (Equation \eqref{fechol}) be zero. A short calculation leads to the following equation for $Q_h$ at the transition temperature:
\begin{multline}
\biggl( 4Q_h + \tfrac{\kappa^2}{4} \tfrac{(1-\alpha Q_h)^2}{1-\beta Q_h} \biggr) \biggl( 6-\tfrac{\kappa^2}{2} \tfrac{1-\alpha Q_h}{1-\beta Q_h} \Bigl[ 2\alpha - \beta \tfrac{1-\alpha Q_h}{1-\beta Q_h} \Bigr] \biggr)^2 \\ - \biggl( 24Q_h + \tfrac{\kappa^2(1-\alpha Q_h)^2}{1-\beta Q_h} - \tfrac{\kappa^2 Q_h}{2} \tfrac{1-\alpha Q_h}{1-\beta Q_h} \Bigl[ 2\alpha - \beta \tfrac{1-\alpha Q_h}{1-\beta Q_h} \Bigr] \biggr)^2 = 0 \; .
\label{Qhtransition}
\end{multline}
Again when $\alpha = \beta = 0$ we recover the usual results \cite{ghsa, wright, belyakov}
\begin{subequations}
\begin{align}
Q_h^{\text{IC}} & = \tfrac{1}{8} \Bigl( 1 - \tfrac{\kappa^2}{3} + \sqrt{1+ \tfrac{\kappa^2}{3}} \Bigr) \; , \\
\tau_{\text{IC}} & = \tfrac{1}{2} \Bigl( 1 - \kappa^2 + \bigl( 1 + \tfrac{\kappa^2}{3} \bigr)^{3/2} \Bigr) \; ,
\end{align}
\end{subequations}
where $\tau_{\text{IC}}$ is the temperature at which the isotropic-cholesteric transition occurs. A situation of some experimental interest occurs when the cholesteric undergoes helix inversion at the isotropic-cholesteric transition temperature \cite{huff}. This case may be solved exactly by setting $Q_h=1/\alpha$ in Equation \eqref{Qhtransition}. We find $\alpha=4$ and
\begin{equation}
\tau_{\text{HI}} = \tau_{\text{IC}} = 1-\kappa^2 \; ,
\end{equation}
which is precisely the transition temperature for a nematic (with non-zero $\kappa$). This particular case provides a guide to the range of values that $\alpha$ can be expected to take in experimental systems. 

\begin{figure}[tb]
\begin{center}
\subfigure[]{
\label{fig:choltrt}
\begin{minipage}[b]{0.48\textwidth}
\centering
\includegraphics[width=65mm]{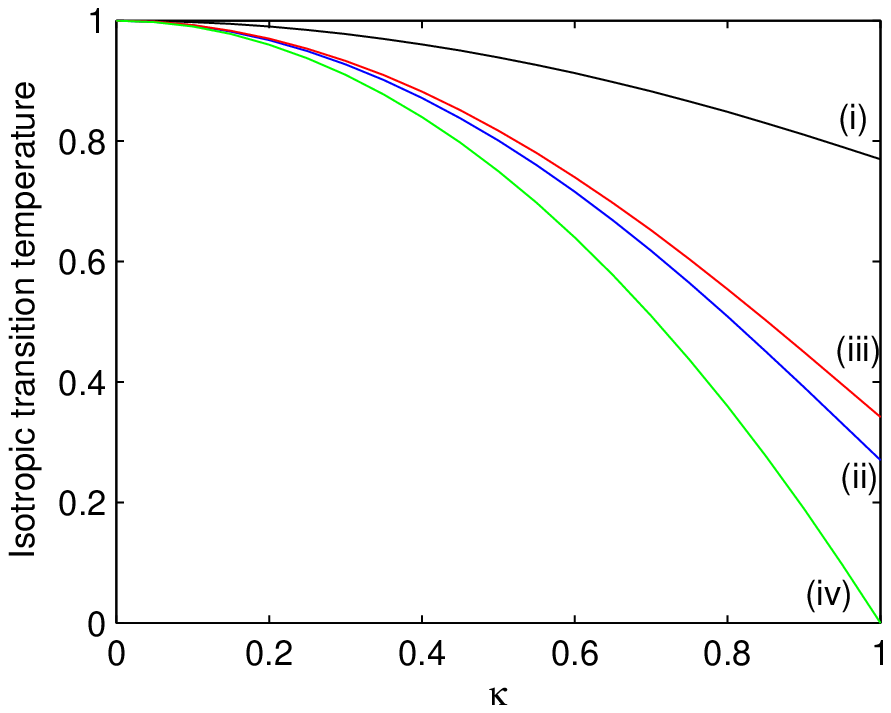}
\end{minipage}} 
\subfigure[]{
\label{fig:cholrw}
\begin{minipage}[b]{0.48\textwidth}
\centering
\includegraphics[width=65mm]{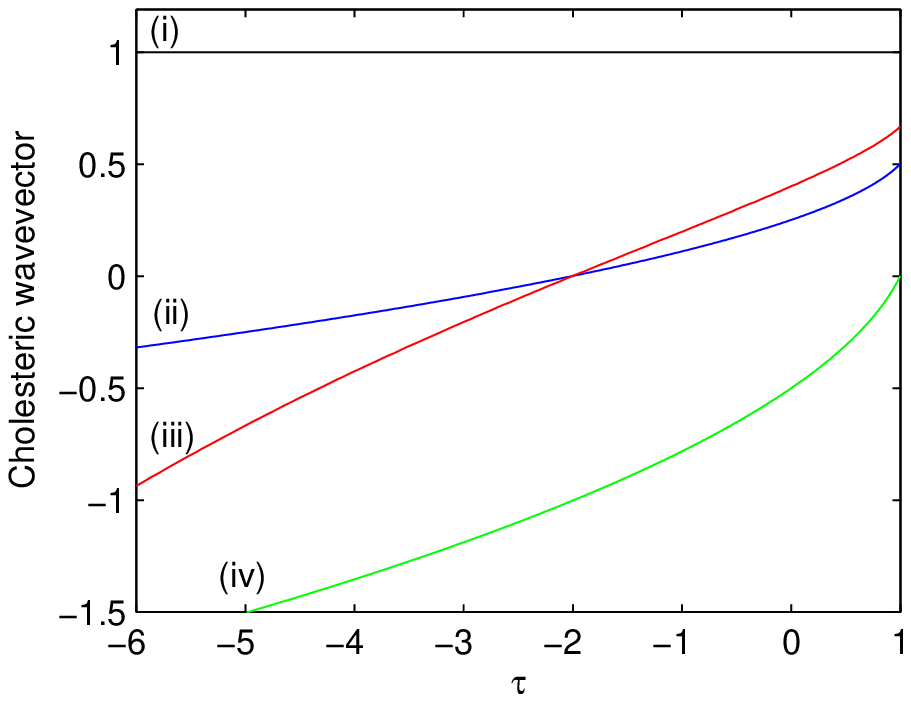}
\end{minipage}}
\end{center}
\caption{(a) Isotropic cholesteric transition temperature as a function of chirality for a range of parameter values: (i) $\alpha=\beta=0$, (ii) $\alpha=2\, ,\, \beta=0$, (iii) $\alpha=2\, ,\, \beta=1$, (iv) $\alpha=4\, ,\, \beta=0$. (b) Temperature dependence of the cholesteric wavevector for the same set of parameters as in (a) (all cases are for $\kappa=0.05$).}
\label{fig:cholplot}
\end{figure}
In the general case the equation can be solved numerically. Figure \ref{fig:choltrt} shows the isotropic-cholesteric transition temperature as a function of chirality for a selection of values of $\alpha$ and $\beta$. The general trend can be easily understood on the basis of the effect that the parameters have on the helical pitch. Increasing $\alpha$ increases the pitch and shifts the liquid crystal towards nematic behaviour, while increasing $\beta$ decreases the pitch and hence shifts away from the nematic state. The temperature dependence of the helical wavevector is shown for the same set of values of $\alpha$ and $\beta$ in Figure \ref{fig:cholrw}. It can be seen that the wavevector varies approximately linearly with temperature except in the immediate vicinity of the transition temperature. It may at first seem unusual that increasing $\beta$ should lead to a larger temperature dependence, however this can be understood by noting that the inversion temperature depends only on $\alpha$ and that increasing $\beta$ leads to a larger wavevector at the transition temperature.

Finally, it is known that if the chirality is sufficiently large then the isotropic-cholesteric transition is second order instead of first. We can calculate the maximum value of $\kappa$ compatible with a first order transition by setting $Q_h=0$ in Equation \eqref{Qhtransition}. The resulting equation is solved to obtain \cite{comment2}
\begin{equation}
\kappa_{\text{max}} = \tfrac{2}{2\alpha-\beta}\Bigl[ -1+\sqrt{1+3(2\alpha-\beta)} \Bigr] \; .
\end{equation}
In what follows it will be seen that this limit is never reached due to the intervention of blue phases.

\section{Numerical Method}
\label{sec:numeric}

Our next aim in this paper is to construct the phase diagram resulting from a minimisation of the modified Landau - de Gennes free energy, Equation \eqref{eq:ldgfreeenergy}, taking into account the two cubic blue phases. Although a certain amount on the blue phases can be done analytically, this approach involves adopting an approximate form for the order parameter and only gives a constrained minimisation. Furthermore, althougth the analytic theory correctly identifies the two structures observed in experiments, it does not reproduce the correct order of appearance of the two phases at low chiralities. A full minimisation can be achieved numerically and has been described recently in Reference \cite{dupuis} for the one elastic constant approximation. The $\mathbf{Q}$-tensor is relaxed towards the minimum of the free energy according to the equation \cite{beris}
\begin{equation}
\frac{\partial Q_{ab}}{\partial t} = \Gamma H_{ab} \; ,
\label{eqQevol}
\end{equation} 
where $\Gamma$ is a collective rotational diffusion constant and the molecular field is given by
\begin{equation}
\mathbf{H} = - \frac{\delta F}{\delta \mathbf{Q}} + \frac{1}{3}\, \text{tr} \Bigl( \frac{\delta F}{\delta \mathbf{Q}} \Bigr) \mathbf{I} \; .
\label{eq:molecularfield}
\end{equation}
Since we are entirely concerned with static equilibrium configurations we have neglected the coupling to fluid flow, both in the above equations and in our numerical simulations. The equations are solved using a three dimensional lattice Boltzmann algorithm, the details of which have been given in Reference \cite{denniston}. 

To study the different phases, cholesteric, BPI or BPII, it is necessary to implement appropriate initial conditions for the simulation. The $\mathbf{Q}$-tensor is initialised using analytic expressions appropriate to the high chirality limit which act to define the symmetry of the chosen phase. Under subsequent numerical evolution according to Equation \eqref{eqQevol} the system relaxes to that structure of the same symmetry which minimises the free energy. We are therefore able to obtain, for any value of the parameters, {\it local} minima of the free energy corresponding to each of the cholesteric and blue phases. The {\it global} free energy minimum was taken to be the smallest of these calculated local minima.

We have seen in Section \ref{sec:cholesteric} that the inclusion of cubic invariants in the gradient free energy leads to a temperature dependent helical pitch in the cholesteric phase, Equation \eqref{kmin}. For the blue phases as well, the unit cell size is temperature dependent, so that to achieve a full minimisation of the free energy it is necessary to set the correct unit cell size in the simulation. This unit cell size is not known {\it a priori}, but rather depends on the magnitude of the order parameter, a quantity which is only determined by the numerical minimisation. Therefore we must introduce a means of determining, and setting, the unit cell size as the $\mathbf{Q}$-tensor evolves during the simulation. We can account for a change in unit cell size by rescaling the gradient contributions to the free energy and molecular field. This is accomplished in practice by changing the elastic constants as follows
\begin{subequations}
\begin{align}
q_0 & = q_0^{\text{init}}/r \; , \\
L_{2a} & = L_{2a}^{\text{init}} \times r^2 \; , \\
L_{3b} & = L_{3b}^{\text{init}} \times r^2 \; ,
\end{align}
\end{subequations}
where $a=1,2$, $b=1,\dots ,9$, a superscript `init' denotes the initial value of a simulation parameter and $r$ is the appropriate rescaling factor, which in previous analytic \cite{ghsa} and numerical \cite{dupuis} work was referred to as the `redshift'. One problem with the analytic theories is that the value of the redshift is not determined exactly, but only for the approximate form of the $\mathbf{Q}$-tensor that is assumed. Similarly, in previous analytic work, the redshift was assumed to take the value suggested by the approximate analytic calculations. The exact redshift for the cholesteric phase could be calculated by obtaining the numerical value of $Q_h$ and using Equation \eqref{kmin}. However, a similar approach is not available for the blue phases and consequently it is more useful (and easier) to calculate it using the free energy as follows: since the free energy is quadratic in gradients, it may be written formally in $\mathbf{k}$-space as
\begin{equation}
F = ak^2 + bk + c \; ,
\end{equation}
where the coefficients $a,b$ and $c$ depend on the $\mathbf{Q}$-tensor, but not on $\mathbf{k}$. The optimum wavevector is given by $k = -b/2a$, and since the coefficients $a$ and $b$ are determined by the simulation every timestep it is straightforward to use these values to determine the exact value for the redshift. 

\begin{figure}[tb]
\centering
\subfigure[]{
\label{fig:hiwavevector}
\begin{minipage}[b]{.48\textwidth}
\centering
\includegraphics[width=65mm]{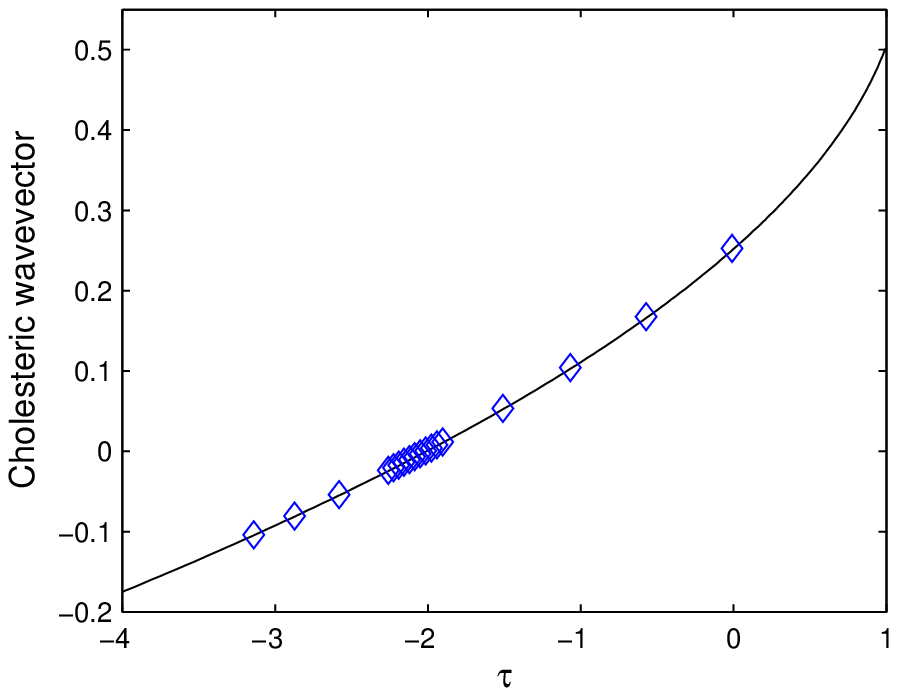}
\end{minipage}} 
\subfigure[]{
\label{fig:hipitch}
\begin{minipage}[b]{.48\textwidth}
\centering
\includegraphics[width=65mm]{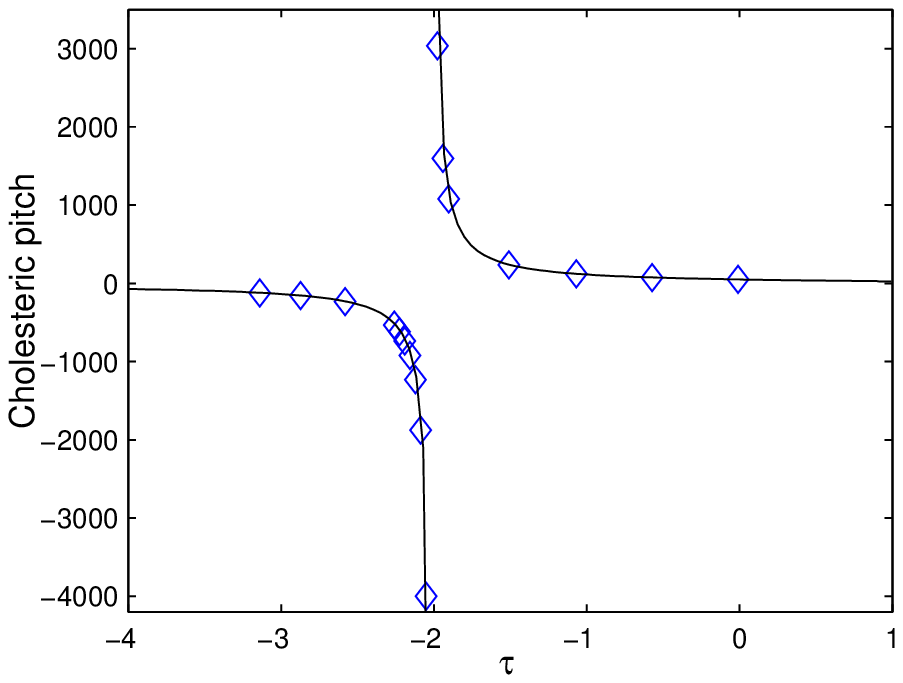}
\end{minipage}}
\caption{Comparison of lattice Boltzmann ($\diamond$) and analytic (solid line) solutions for the helical pitch in a cholesteric displaying helical sense inversion. (a) Cholesteric wavevector and (b) pitch against reduced temperature, $\tau$.}
\label{fig:hi}
\end{figure}
In order to verify that the procedure was working successfully, and to check the level of accuracy that could be obtained, we used the lattice Boltzmann algorithm to calculate the wavevector of a cholesteric undergoing helix inversion and compared it to the theory described in Section \ref{sec:cholesteric}. Simulation parameters were chosen to set $\alpha = 2.0$ and $\kappa \approx 0.096$. The results are shown in Figure \ref{fig:hi}. We have plotted both the helical wavevector, which is the relevant quantity theoretically, and the pitch, since this is more frequently given in experimental work. As can be seen the agreement is excellent even very close to the inversion point.

\subsection{Results}
\label{sec:numericresults}

\begin{figure}[tb]
\begin{center}
\subfigure[]{
\label{fig:pdkmin}
\begin{minipage}[b]{0.48\textwidth}
\centering
\includegraphics[width=65mm]{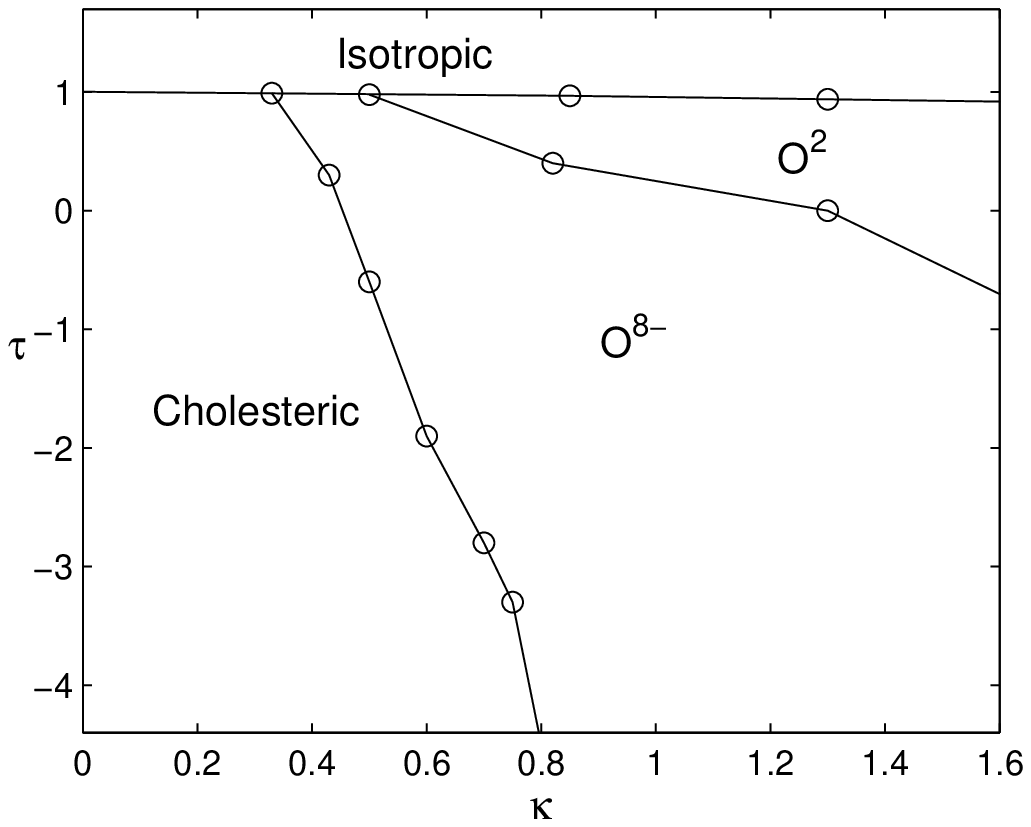}
\end{minipage}} 
\subfigure[]{
\label{fig:alexpd}
\begin{minipage}[b]{0.48\textwidth}
\centering
\includegraphics[width=65mm]{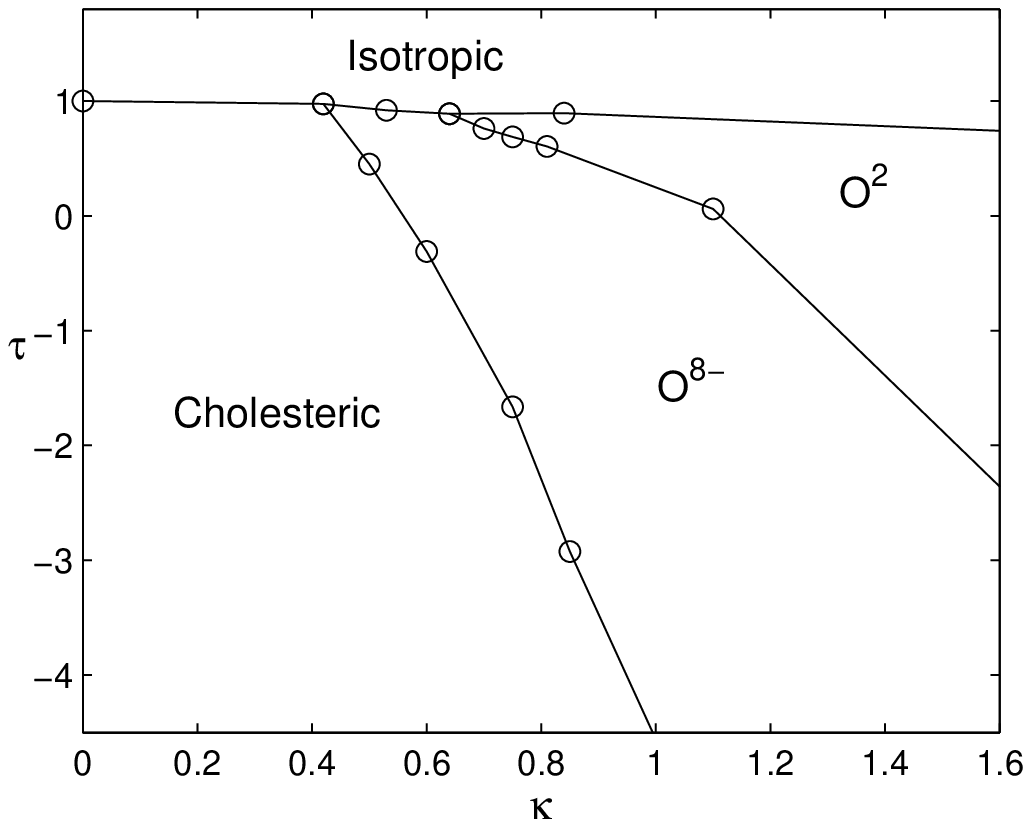}
\end{minipage}}
\end{center}
\caption{Numerical phase diagrams in the one elastic constant approximation. $\tau$, the reduced temperature, and $\kappa$, the chirality, are defined by Equations \eqref{eq:deftaukappa}. (a) Optimisation of the unit cell size. (b) Phase diagram with fixed unit cell size, after \cite{dupuis}.}
\label{fig:kmin}
\end{figure}

In this section we describe the numerical phase diagram obtained using the lattice Boltzmann algorithm outlined above. Figure \ref{fig:pdkmin} shows the phase diagram obtained for the one elastic constant approximation with numerical optimisation of the unit cell size. For comparison, the phase diagram for a fixed unit cell size determined in Reference \cite{dupuis} is reproduced in Figure \ref{fig:alexpd}.
We note that optimisation of the unit cell size has extended the range of stability of BPI, both relative to the cholesteric phase and relative to BPII. The movement of the cholesteric phase boundary is quite significant, with the triple point moving to lower chirality by about 20\%. This is due to the optimum redshift taking a lower value at these chiralities than was assumed previously. Based on analytic calculations of Grebel {\it et. al.} \cite{ghsb} the redshift was assigned the value 0.79 in \cite{dupuis}. However, we find a much lower value, with average 0.68 at the cholesteric-BPI phase boundary (the value is roughly independent of temperature, except very close to the isotropic transition temperature). In contrast, at the BPI-BPII phase boundary we obtain an average redshift of 0.77 for BPI and 0.86 for BPII. These values are in better agreement with the analytic results of Grebel {\it et. al.}, primarily because the chirality at this phase boundary is much closer to the values of chirality where the analytic calculations predict the blue phases are stable. In what follows we will use Figure \ref{fig:pdkmin} as a reference phase diagram relative to which the effect of varying the Landau - de Gennes parameters can by measured.

\begin{figure}[tb]
\begin{center}
\subfigure[]{
\label{fig:pdl3}
\begin{minipage}[b]{0.48\textwidth}
\centering
\includegraphics[width=65mm]{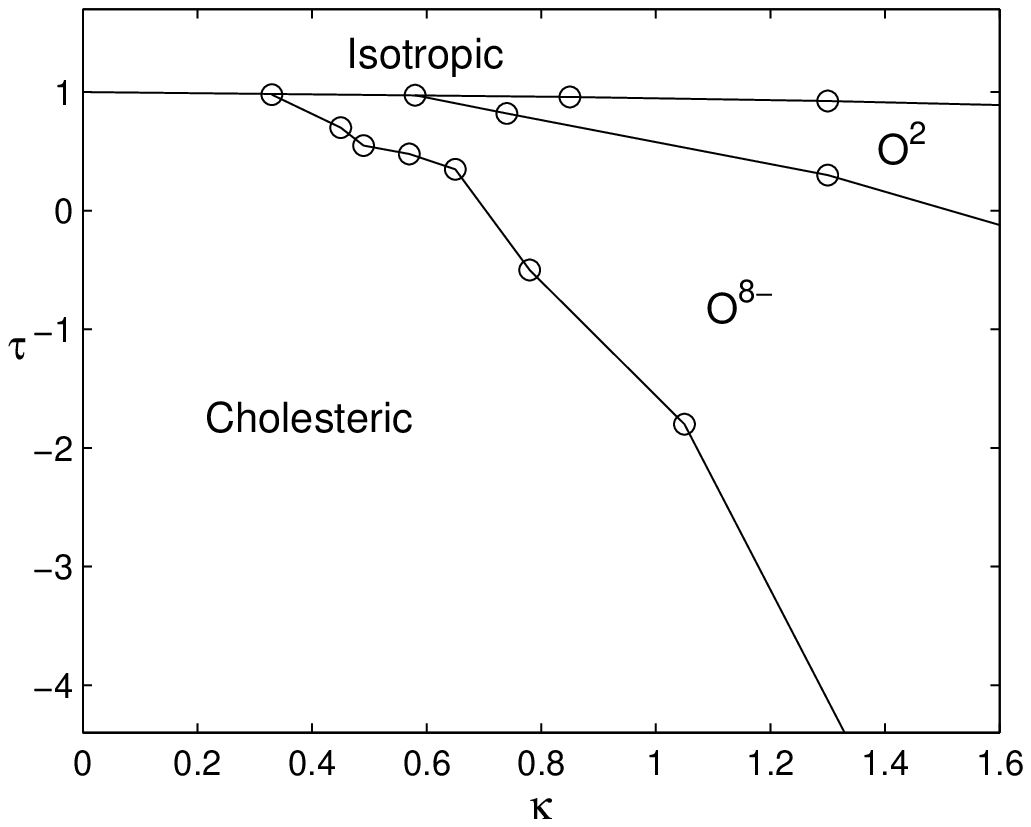}
\end{minipage}} 
\subfigure[]{
\label{fig:l2pd}
\begin{minipage}[b]{0.48\textwidth}
\centering
\includegraphics[width=65mm]{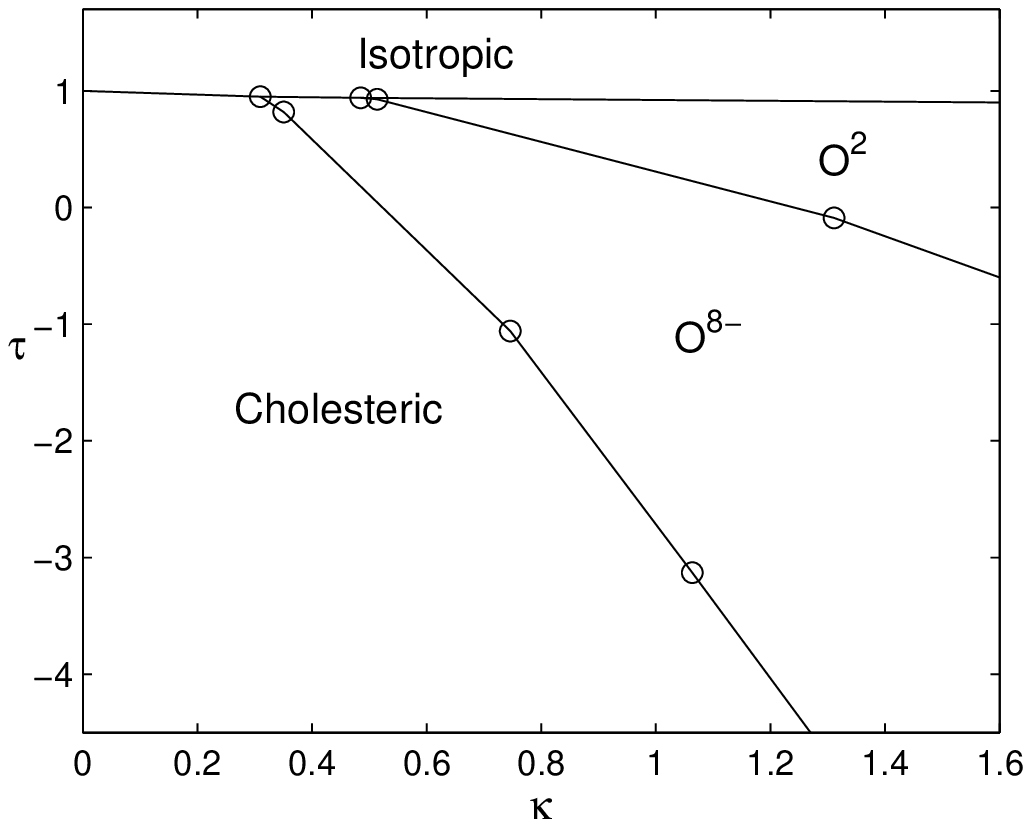}
\end{minipage}}
\end{center}
\caption{Phase diagrams obtained for different values of the elastic constants. $\tau$, the reduced temperature, and $\kappa$, the chirality, are defined by Equations \eqref{eq:deftaukappa}. (a) $K_1^{\text{F}}=K_2^{\text{F}}=0.5 K_3^{\text{F}}$, (b) $K_1^{\text{F}}=K_3^{\text{F}}=1.5 K_2^{\text{F}}$.}
\label{fig:elcons}
\end{figure}

We next constructed phase diagrams for different ratios of the elastic constants. Here one expects that the qualitative features of the phase diagram will be retained, but it is nonetheless of interest to determine how large a quantitative shift can be obtained. Figure \ref{fig:elcons} shows the phase diagrams obtained upon separately varying the twist and bend elastic constants.
To investigate the effect of the bend elastic constant we chose parameter values $L_{21}=L_{22}=L_{34}=0.02, \, L_{38}=0$ which corresponds to a ratio of bend to splay of about 1.75, while splay and twist remain degenerate. The stability of BPI is seen to decrease quite significantly relative to the cholesteric phase while at the same time there is a small increase in stability over BPII. There is only a minor shift in the cholesteric-BPI phase boundary at the transition temperature, however, as the temperature decreases the shift becomes larger. For example at a reduced temperature of $\tau = -2$ the phase boundary occurs at a chirality of $\kappa \approx 1$, representing a shift to higher chiralities of almost 70\% as compared to the one elastic constant case. The value of the redshift is markedly increased, with the average value for BPI at the cholesteric boundary being 0.81. However, it was shown in Section \ref{sec:cholesteric} that the cholesteric wavevector depends on the value of $L_{34}$ and hence has also increased. This raw value of the redshift no longer represents the most directly accessible quantity and a more relevant figure is the ratio of the BPI redshift to that of the cholesteric, since this can be measured experimentally by means of the discontinuity of the back-scattered Bragg peak. For this quantity we obtain an average of 0.67 and note the strong similarity of this value with that obtained in the one elastic constant approximation.

At the BPI-BPII phase boundary the redshift is 0.87 and 0.97 for BPI and BPII respectively. Again, it is the ratio of these values which is more directly relevant to experiment. The ratio of the $O^{8-}$ unit cell size to that of $O^2$ is 0.90, and again we note a strong similarity with the ratio obtained from the one elastic constant approximation, 0.89.

The value of the twist elastic constant is controlled by the Landau - de Gennes parameter $L_{22}$. In most liquid crystals the twist elastic constant is smaller than either splay or bend. In order to match this, we constructed the phase diagram for parameter values $L_{21}=0.02, \, L_{22}=0.04, \, L_{34}=L_{38}=0$, which is shown in Figure \ref{fig:l2pd}. This choice of parameters resulted in a ratio of splay to twist of about 1.5, while splay and bend remained degenerate. Again we observe that the stability of BPI is reduced relative to the cholesteric phase by an amount similar to that seen by varying the bend elastic constant. The values of the redshift for both BPI and BPII are only very slightly increased relative to their values in the one elastic constant limit, while the cholesteric wavevector is insensitive to the value of $L_{22}$. This reveals an intriguing feature, that while the phase boundaries and absolute values of the redshift can vary appreciably, the ratios of the redshift for the different phases, and hence the discontinuities in back-scattered Bragg peaks, are essentially independent of the values of the elastic constants. 

\begin{figure}[tb]
\begin{center}
\subfigure[]{
\label{fig:bppdchiral}
\begin{minipage}[b]{0.48\textwidth}
\centering
\includegraphics[width=65mm]{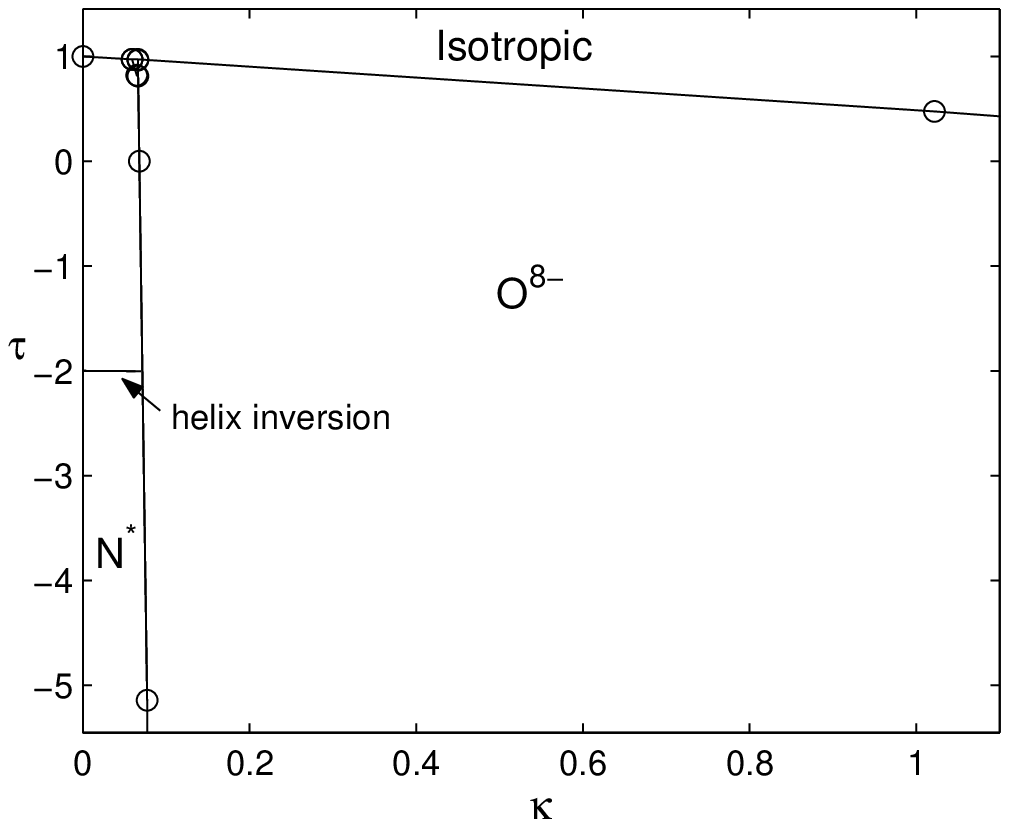}
\end{minipage}} 
\subfigure[]{
\label{fig:bppdchiral2}
\begin{minipage}[b]{0.48\textwidth}
\centering
\includegraphics[width=65mm]{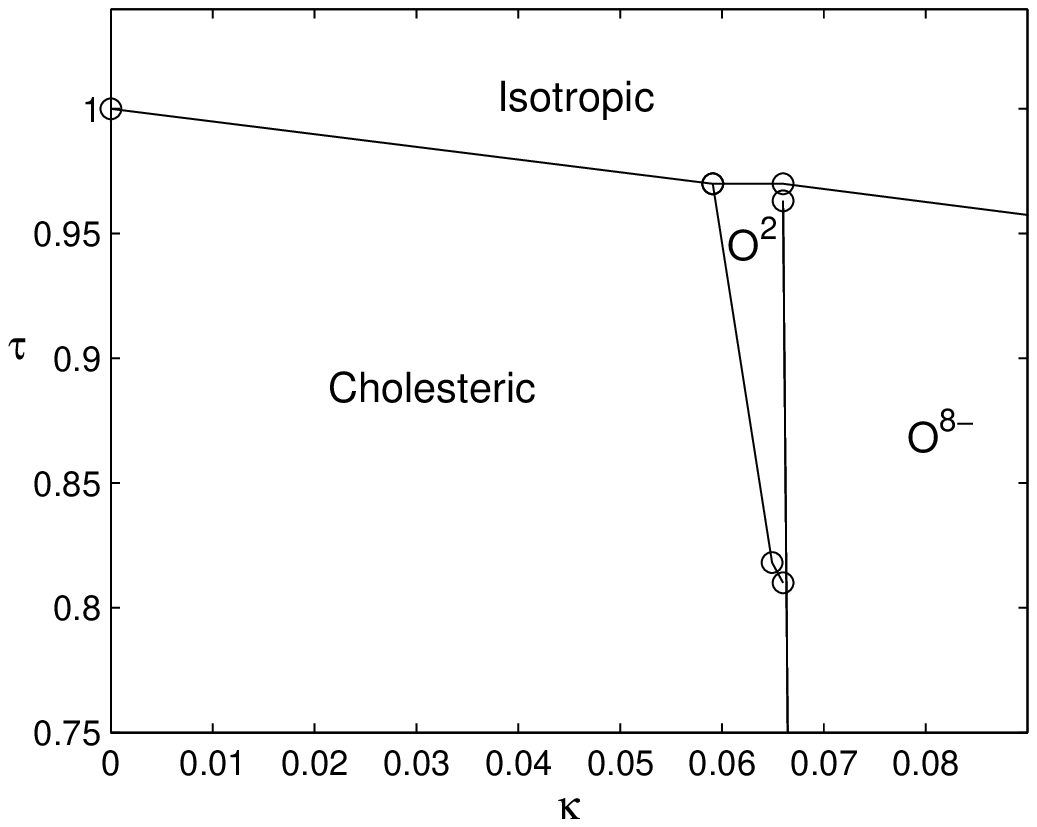}
\end{minipage}}
\end{center}
\caption{(a) The numerical phase diagram obtained with the chiral invariant, \eqref{eq:cubicchiral}, added to the free energy. The magnitude of this term was chosen so as to produce helical sense inversion in the cholesteric phase at a temperature not far below the isotropic transition temperature. $\tau$, the reduced temperature, and $\kappa$, the chirality, are defined by Equations \eqref{eq:deftaukappa}. (b) An enlargement of the region near the isotropic transition temperature. Note the reversal in the order of appearance of BPI and BPII as a function of chirality.}
\label{fig:stable}
\end{figure}

Finally, we investigated the effect of the chiral cubic invariant on the blue phases. We chose to use $L_{21}=L_{22}=L_{34}=0.02, \, L_{38}=0.04, \, \alpha=2.0$. (For the cholesteric phase this sets $\beta=0$ and gives a ratio of bend to splay of about 1.6.) The phase diagram is shown in Figure \ref{fig:stable}.
The value of $\alpha$ is such that the cholesteric undergoes helical sense inversion at a reduced temperature of about $\tau=-2$. What is remarkable is the dramatic increase in stability of BPI relative to the cholesteric phase. The region of stability has been increased down to chiralities as low as $\kappa=0.07$ and at such low chiralities the phase boundary is essentially independent of $\kappa$ for all $\tau$. In addition, we find a very small region of stability for BPII located close to the isotropic transition. As shown in greater detail in Figure \ref{fig:bppdchiral2} this occurs in a narrow temperature interval at chiralities lower than those for which BPI is stable, representing a reversal of the order of appearance of the two blue phases. In the region where BPII is stable we find a redshift of 0.36. In contrast to the situation without the chiral invariant, this is {\it smaller} than the BPI redshift, which takes the value 0.43. Again, the absolute values of the blue phase redshifts are not as relevant as their ratios to that of the cholesteric phase. In this case we find the cholesteric redshift is 0.49 at the isotropic transition temperature, giving ratios of 0.88 for BPI and 0.74 for BPII, both of which are significantly different to those obtained in the one elastic constant limit.

\begin{figure}[tb]
\begin{center}
\subfigure[]{
\label{fig:bpredtemp}
\begin{minipage}[b]{0.48\textwidth}
\centering
\includegraphics[width=65mm]{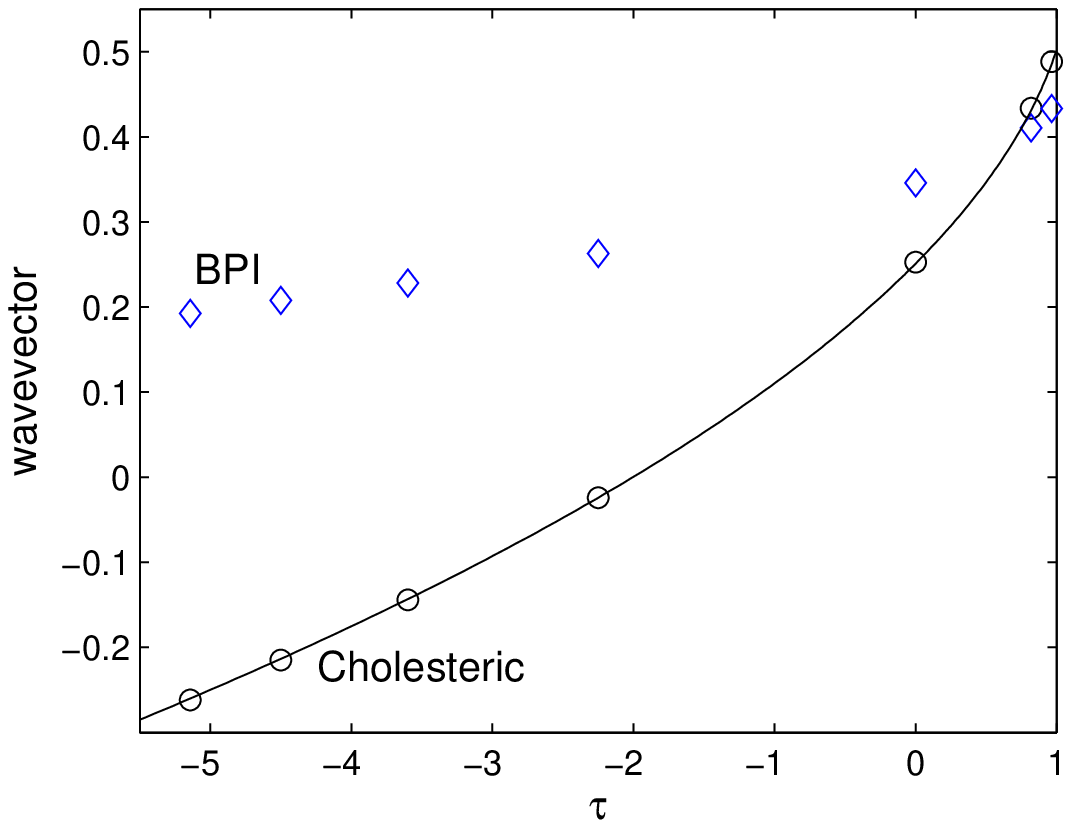}
\end{minipage}} 
\subfigure[]{
\label{fig:bppitchtemp}
\begin{minipage}[b]{0.48\textwidth}
\centering
\includegraphics[width=65mm]{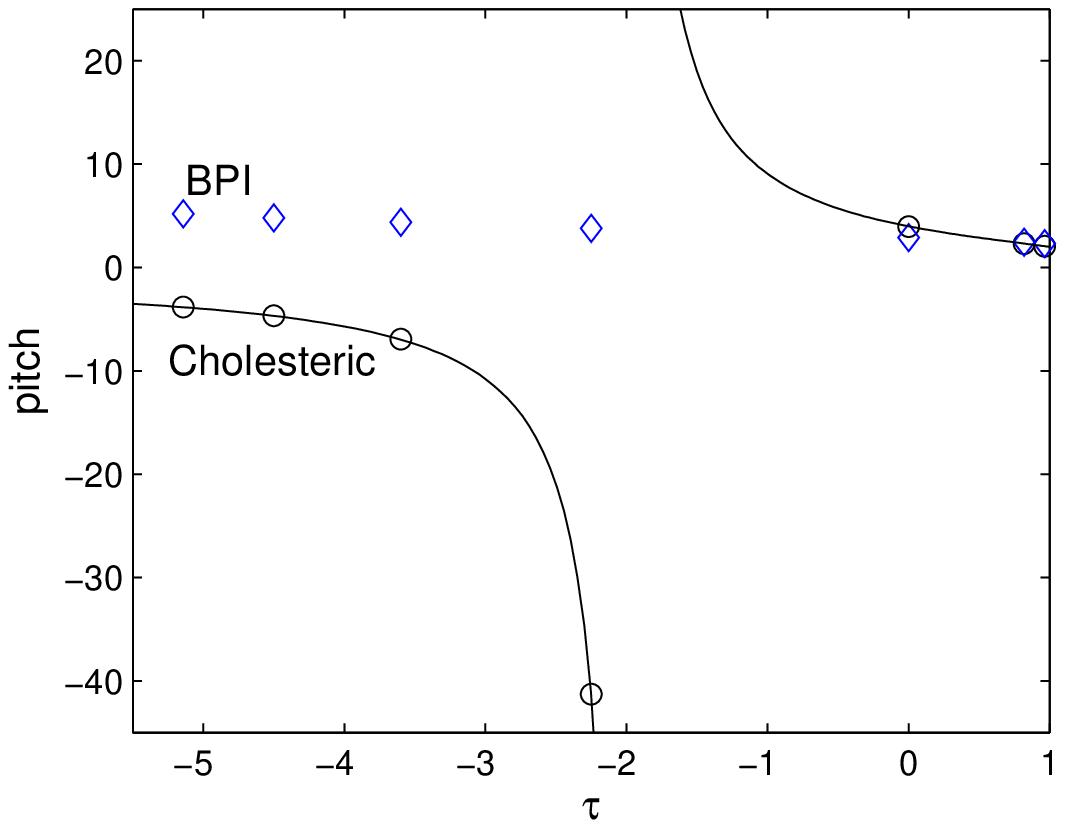}
\end{minipage}}
\end{center}
\caption{Lattice Boltzmann results for the temperature dependence of the (a) wavevector and (b) pitch of the unit cell of BPI along the cholesteric-BPI phase boundary at $\kappa \approx 0.07$ \hspace{3mm} $( \diamond ) \;$. For comparison the cholesteric wavevector and pitch are also given: lattice Boltzmann results $\; ( \circ ) \;$, results from the calculation presented in Section \ref{sec:cholesteric} (solid line).}
\label{fig:bptemp}
\end{figure}

Since BPI is now stable over a much larger temperature range it displays a significant variation in unit cell size as the temperature is lowered. As an illustration of this, the BPI redshift is 0.19 at a reduced temperature of $\tau \approx -5$, corresponding to more than a two-fold increase in the lattice parameter. A plot of the temperature dependence of the BPI redshift is shown in Figure \ref{fig:bptemp}. For comparison the helical wavevector of the cholesteric phase is also plotted on the same graph.

\section{Conclusion}
\label{sec:conclusion}

We have investigated numerically the phase diagram of the cholesteric blue phases for a range of parameter values within the framework of a modified Landau - de Gennes theory. The traditional Landau - de Gennes theory has long been known to only accommodate two independent Frank elastic constants and to have a temperature independent cholesteric pitch. Both of these shortcomings were overcome by retaining terms of cubic order in the $\mathbf{Q}$-tensor in the expansion for the gradient free energy. Since the new terms were added specifically to remove the degeneracy between splay and bend, and to give a temperature dependence to the helical pitch they possess a clear and simple physical interpretation. In particular the value of the parameter $\alpha$ which controls the strength of the chiral cubic invariant should be relatively easy to estimate on the basis of Equation \eqref{hitemp} for the helix inversion temperature. The magnitude of the elastic constants for the achiral cubic invariants is more difficult to ascertain. Although this may be estimated from the ratios of the Frank elastic constants it is clear that because there are more Landau - de Gennes elastic constants than Frank, the latter are insufficient to uniquely determine the former (an estimation of the magnitude of the cubic Landau - de Gennes elastic constants was given in \cite{berreman}).

The modified Landau - de Gennes theory that we have investigated provides a phenomenological description of helical sense inversion in the cholesteric phase. The inversion arises as a natural consequence of the presence of including higher order chiral invariants in the gradient free energy. Although such higher order contributions are usually neglected since they are deemed small compared to the terms already retained, the fact that helix inversion is observed experimentally demonstrates that these terms can play a significant role. Mathematically, we comment that for systems undergoing first order phase transitions (such as liquid crystals), although the order parameter is small it is not infinitesimal, and therefore the relative magnitude of a given term in the Landau expansion depends not only on the power of the order parameter but also on the size of any numerical coefficient premultiplying it.

The primary aim of this paper was to investigate how much the properties of the blue phases could be changed within the framework of Landau - de Gennes theory. In this regard we have shown that the retention of cubic order terms in the gradient free energy can lead to considerable changes in the size of the blue phase unit cell and in their phase diagram. Most dramatic amongst the results is the increase in stability of blue phase I obtained in systems where the cholesteric undergoes helical sense inversion. Again we comment that this significant, qualitative change in the phase diagram arises from retaining cubic order terms and demonstrates that these give rise to more than just small changes in the physical properties. 

It is of interest to consider whether the mechanism considered here is a candidate to account for the increased range of stability in blue phase I recently reported by Coles and Pivnenko \cite{coles}. It seems not, as apart from the large temperature range most features of their blue phase differ from those obtained for the choice of parameters we made here. For example, the numerics show an increase in the size of the BPI unit cell with decreasing temperature, while in the experimental system the unit cell size shows a small decrease. Also, numerically we find that BPII has a larger unit cell than BPI at the transition between the two, in contrast to what is found experimentally. However, we remark that the parameter space in the Landau - de Gennes theory is large enough that these discrepancies could well be resolved by a different choice of parameters.


\begin{thebibliography}{99}
\bibitem{comment1} This equation is exact in that the inclusion of all other cubic invariants in the gradient free energy simply alters the value of $\alpha$, but makes no other change.
\bibitem{comment2} The solution is identified by the requirement that it approach the correct value, $\kappa=3$, as $(2\alpha-\beta) \rightarrow 0$.
\end{thebibliography}
\end{document}